\documentclass[10pt,twocolumn]{article}

\usepackage{relsize}
\usepackage{multicol}
\usepackage{graphicx}
\usepackage{amsmath}
\usepackage{lipsum}
\usepackage{color}
\usepackage{siunitx}
\usepackage{authblk}
\usepackage{subfigure}
\usepackage[margin=2cm]{geometry}

\title{Drop morphologies on flexible fibers: influence of elastocapillary effects}
\author[1,2]{Alban Sauret}
\author[3]{Fran\c{c}ois Boulogne}
\author[2,3]{Katarzyna Somszor} 
\author[2]{\\ Emilie Dressaire}
\author[3]{Howard A. Stone}
\affil[1]{Surface du Verre et Interfaces, UMR 125 CNRS/Saint-Gobain, 93303 Aubervilliers, France (E-mail: alban.sauret@saint-gobain.com)}
\affil[2]{Department of Mechanical and Aerospace Engineering, New York University Tandon School of Engineering, Brooklyn, NY 11201, USA}
\affil[3]{Department of Mechanical and Aerospace Engineering, Princeton University, Princeton, New Jersey 08544, USA}
\date{ }
\begin{document}

\twocolumn[
    \begin{@twocolumnfalse}
        \maketitle
        \begin{abstract}
            Various materials are made of long thin fibers that are randomly oriented to form a complex network in which drops of wetting liquid tend to accumulate at the nodes. The capillary force exerted by the liquid can bend flexible fibers, which in turn influences the morphology adopted by the liquid. In this paper, we investigate, the role of the fiber flexibility on the shape of a small volume of liquid on a pair of crossed flexible fibers, through a model situation. We characterize the liquid morphologies as we vary the volume of liquid, the angle between the fibers, and the length of the fibers. The drop morphologies previously reported for rigid crossed fibers, \textit{i.e.}, a drop, a column and a mixed morphology, are also observed on flexible crossed fibers with modified domains of existence. In addition, at small tilting angles between the fibers, a new behavior is observed: the fibers bend and collapse. Depending on the volume of liquid, a thin column with or without a drop is reported on the collapsed fibers. Our study suggests that the fiber flexibility adds a rich variety of behaviors that may be important for some applications. \\
            \medskip\medskip\medskip
        \end{abstract}
    \end{@twocolumnfalse}
]

\section{Introduction}

A capillary bridge between two spheres or two surfaces is known to exert a force that tends to keep the two solid bodies together \cite{FISHER1981528,Mastrangelo:1993uf,Willett:2000ic}. For soft surfaces, the combination of the capillary force and the bending and/or stretching of the material leads to an elastocapillary equilibrium in which the elastic deformations of the substrate is related to the intensity of the capillary adhesion \cite{Roman:2010ipa,Liu:2012gk}. Elastocapillary effects are, for instance, responsible for the formation of bundles with flexible beams \cite{Bico:2004vi,Singh:2014bk,Pokroy:2009ip} or fibers \cite{Boudaoud:2007hf,Wang:2014ej}, the collapse of beams in MEMS applications \cite{Pokroy:2009ip,Chandra:2009jc,Chandra:2010eo,DeVolder:2013cs} and in carbon nanotube \cite{Chakrapani:2004up} or in biological situations such as the wetting of feathers \cite{Hartung:1967tm}.

Elastocapillary effects can be particularly important when considering thin elongated structures, such as fibers, whose length can be several order of magnitude larger then their diameter; such fibers are therefore prone to bend \cite{Boudaoud:2007hf,Py2007a,Duprat:2012kya,elettro2016drop,schulman2016elastocapillary}. In addition, fibers, rigid or flexible, can be found in a variety of applications such as in wet hair \cite{Bico:2004vi}, in the textile industry \cite{Minor:1959wa,Miller:1967vy,Eadie:2011hm}, in filters \cite{Mullins:2004es,Contal:2004bg}, fog-harvesting nets \cite{Ju:2012gl,Park:2013go,LeBoeuf:2014hu} or insulation materials such as glass wool \cite{Sauret:2015ch,Bintein:2015tw}. Therefore, many studies have considered the addition of liquid on an array of fibers through model systems that consist of a finite volume of wetting liquid deposited on a single or a pair of fibers \cite{Princen:1970ts,Carroll:1986uia,Quere:1999wv}. These model situations allow investigation of the equilibrium shapes of the liquid and their influence on the drying process \cite{Sutter:2010jf,Boulogne:2015kl}, the capture of impacting drops \cite{Lorenceau:2004da}, the condensation of liquid \cite{Zhang:2015dt} or the motion of drops along fibers \cite{gilet2009digital,Gilet:2010ti,Boulogne2012,weyer2015compound,Blanchette:2016ds}.

Past studies have considered the morphology adopted by a liquid drop deposited on a single fiber. For example, the addition of liquid on a single fiber can come from coating methods \cite{Quere:1999wv}, the condensation of liquid \cite{Zhang:2015dt} or the impact of small droplets that are captured below an impact velocity threshold \cite{Lorenceau:2004da,Piroird:2009hu,Dressaire:2015hp}. Once on the fiber, the drop can either adopt an axisymmetric (barrel shape) morphology or an asymmetric (clamshell) morphology where the drop sits on one side of the fibers \cite{Carroll:1976uj,Carroll:1986uia,McHale:2001vy,McHale:2002wu,Wu:2014eb}.

Considering rigid fibers, the presence of a second fiber close to the first one leads to different morphologies that depend on the contact angle of the liquid with the fiber, the inter-fiber distance, the tilting angle between them and the volume of liquid that is deposited. For instance, for two parallel fibers separated by a distance $2\,d$, the seminal work of Princen \cite{Princen:1970ts} shows that the liquid can either spread in a long liquid column, whose width is of the order of the fiber diameter if the fibers are close enough, or adopt a more compact drop shape when the inter-fiber distance is increased. More recently, three different possible morphologies have been reported for fibers of radius $a$ randomly oriented, i.e., tilted with an angle $\delta$ and separated by a minimum distance $h$. Depending on $\delta$, $\tilde{h}=h/a$, and the dimensionless volume of liquid $\tilde{V}=V/a^3$, the liquid can be in a column shape, a mixed morphology state where a drop lies at the end of a column, or a drop centered at the node, {\textit{i.e.}}, the point where the two fibers are in contact \cite{Sauret:2014fb,Sauret:2015ba,Sauret:2015ch}.

However, some fibrous materials, involve fibers that are very thin and long. Such high aspect ratio fibers are typically flexible. Therefore, the presence of liquid forming capillary bridges between fibers is susceptible to modify the liquid morphologies and the organization of the fibers in a randomly oriented fibers array. Indeed, the presence of liquid can lead to clustering of flexible structures as shown by Py \textit{et al.} \cite{Py2007a} who investigated the behavior of a model brush of parallel flexible fibers withdrawn from a liquid bath. 
To describe their experimental findings, the authors defined an elastocapillary lengthscale \cite{Bico:2004vi,Roman:2010ipa}:
\begin{equation}\label{bico}
  \mathlarger{\ell}_{EC}=\sqrt{\frac{E\,I}{\gamma\,a}},
\end{equation}
where $E$ is the Young's modulus of the fiber, $I=\pi\,a^4/4$ is the polar moment of inertia, $a$ the radius and $\gamma$ the liquid surface tension. 
{The elastocapillary length can be derived from the balance between the capillary energy $\gamma \, a \, L_{fib}$ and the bending energy $E\,I\,L_{fib}/R^2$ where $R$ is the typical radius of curvature.
This balances leads to $\ell_{EC} = R$, and this elastocapillary length can be seen as a local radius of curvature of the fiber.}

Because the capillary force and the elasticity of the fibers can lead to the deformation of the fibers \cite{soleimani2015capillary}, different configurations have been investigated. In particular, Duprat {\textit{et al.} considered parallel fibers whose spacing and length are controlled and a known volume of liquid is deposited} \cite{Duprat:2012kya}, Different morphologies were observed: a drop state, a column morphology or a partially wetting morphology. In addition, the presence of liquid was shown to bend the fibers and influence the final liquid morphology. In their configuration, the elastocapillary length is around $\ell_{EC}\sim 55\,{\rm cm}$ whereas the length of the fibers $L_{fib}$ is typically a few centimeters. In such model experiments, although $L_{fib} \ll \ell_{EC}$, the effects of the flexility can be observed experimentally leading to the presence of new liquid morphologies.

However, the more general situation of tilted flexible fibers, encountered in arrays of fibers, has not been studied. The presence of liquid at a node is expected to introduce a capillary force and a torque that can modify the orientation of the fibers and lead to different morphologies \cite{Claussen2011}. However, no systematic experimental investigations has been reported to date. To investigate this situation, we consider the situation of two crossed fibers, clamped at one of their ends and free to move at the other, in contact at the node and tilted with an angle $\delta =  [0,\,90]^{\rm o}$. Using experiments, we then investigate the morphologies and the equilibrium orientations of the fibers while varying the imposed initial tilting angle $\delta$ and the volume of liquid $V$ deposited on the fibers. We are then able to highlight the presence of two new morphologies compared to the situation of crossed rigid fibers: a column collapsed state and a drop-collapsed state. We first present the experimental method in section 2. In section 3, we describe the new morphologies and a morphology diagram in the ($\delta$, $\tilde{V}$) space is reported. We then rationalize our results to explain these new morphologies. Finally, we highlight in section 4 that hysteresis effects can be large in such systems owing to capillary effects.

\section{Experimental protocol}

\begin{figure}
    \centering
\includegraphics[width=8cm]{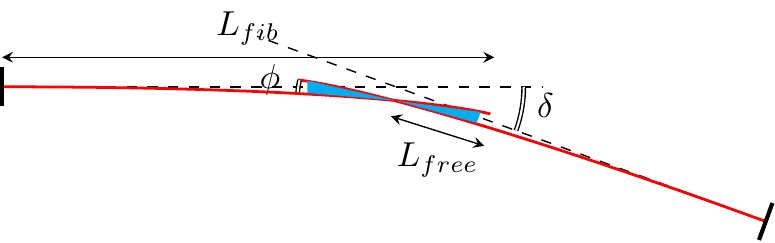}
\caption{Schematic of the experimental setup showing the wetting liquid and the tilted fiber. Both fibers are clamped at one end and free at the other end such that they can bend owing to a capillary force.
    Experimentally, we impose a tilt angle $\delta$ and the elasticity of the fibers leads to a new angle $\phi$.
    }\label{fig:schema}
\end{figure}

In our experiments, the fibers consist of thin capillary cylindrical tubes (purchased from VitroCom) of outer diameter $2\,a=250\,\mu{\rm m}$ and inner diameter $2\,b=150\,\mu{\rm m}$.
We seal the end of the tube with a small amount of epoxy glue so that no liquid can flow into the capillary tube.
We have measured the bending modulus of these fibers using their deflection under their own weight and obtained $B=8.1\times 10^{-6}\,{\rm Pa \cdot m^4}$, in agreement with the estimated theoretical value $B = E I = \pi\, E (a^4 - b^4) / 4$, where $I$ is the polar moment of inertia of the cross section of the fiber.
The weight per unit length is $\rho_l=3.5\times 10^{-5} {\rm kg \cdot m^{-1}}$. Therefore, the value of the elastocapillary length in our system is $\ell_{EC} \simeq 1.8\,{\rm m}$, which is of the same order of magnitude as the one of previous experiments investigating elastocapillary effects between parallel fibers \cite{Duprat:2012kya}.
{
Thus, in this paper, we study a system in which the typical drop size $V^{1/3}$ is much smaller than the elastocapillary length, \textit{i.e.}, $V^{1/3}\ll \ell_{EC}$.
As a consequence the fibers cannot bend on the lengthscale of the drop.
Roman and Bico studied the opposite limit and they observed that a soft fiber coils around a liquid drop \cite{Roman:2010ipa}.
}

One extremity of each fiber is clamped on a ($x,y,z$) microcontroller linear stage (PT3, Thorlabs).
Fibers cross at the node.
One of the controllers is allowed to rotate around the node with a rotation mount (PR01, Thorlabs) to accurately tune the angle $\delta$ between the fibers.
We denote $L_{fib}$ the total length of a fiber and $L_{free}$ the distance between the node and the tip of the fiber (Fig. \ref{fig:schema}).
In all experiments presented in this paper, the lengths $L_{fib}$ and $L_{free}$ are identical for both fibers.
The vertical deflection of fibers due to their own weight and the weight of the drop is limited to $3$ mm for the largest fiber length ($L_{fib}=90$ mm) and the largest drop volume ($V = 8 \mu\ell$). We should emphasize that we could not use longer fibers  since gravity effects will then become too important.

Before each experiment, the horizontal positions of the microcontrollers are adjusted to ensure that $L_{free}$ remains constant while varying the tilting angle $\delta$.
The vertical positions of the linear microcontrollers are then adjusted to bring the fibers into contact.
In all the experiments, we use a perfectly wetting fluid ($5$ cSt silicone oil, contact angle $\theta = 0^{\circ}$, density $\rho=918\,{\rm kg\cdot m^{-3}}$, surface tension $\gamma=19.7\,{\rm mN \cdot m^{-1}}$).
A volume $V\in[1,\,8]\,\mu\ell$ is initially dispensed with a micropipette (Eppendorf, Research Plus) at the node of the fibers.
Because the fibers are flexible and bend under capillary effects, the apparent angle between the fiber $\phi$ can be smaller than the tilting angle without liquid $\delta$ as illustrated in the schematic Fig. \ref{fig:schema}.
In our experiments, we assume that the gravitational effects on the liquid bridge can be neglected.
Indeed, the capillary length, $\ell_c=\sqrt{\gamma/(\rho\,g)} \simeq 1.5\,{\rm mm}$, is larger than the typical height $H$ of the liquid.
In addition, we neglect gravitational effects on the fibers because, in the absence of liquid, the fibers do not bend significantly under their own weight with the chosen length $L_{fib}$, as noted above.

We start the experiment at a large tilting angle between the fibers (typically $\delta=60^\circ$) and the angle is reduced to the desired value.
The imposed tilting angle $\delta$ is directly measured on the rotation microcontroller with an uncertainty of about $0.5^{\circ}$.
We wait for a few minutes to ensure that the final steady liquid morphology is reached and record images from the top and from the side using Nikon cameras (D5100 and D7100) and 105 mm macro objectives.
The induced angle $\phi$ is measured using the top view images (corresponding to the schematic in Fig. \ref{fig:schema}) by image processing with an uncertainty of about $0.5^{\circ}$.
When a steady morphology has been reached and the images recorded, we increase or decrease the angle $\delta$ between the fibers and ensure that $L_{free}$ remains constant and wait until a new steady state is reached. Using this method, we explore the influence of the volume of liquid, the tilting angle $\delta$ between the fibers, the total length $L_{fib}$ and the free end $L_{free}$ of the fibers on the final morphology.

\begin{figure}
\begin{center}
\includegraphics[width=8.5cm]{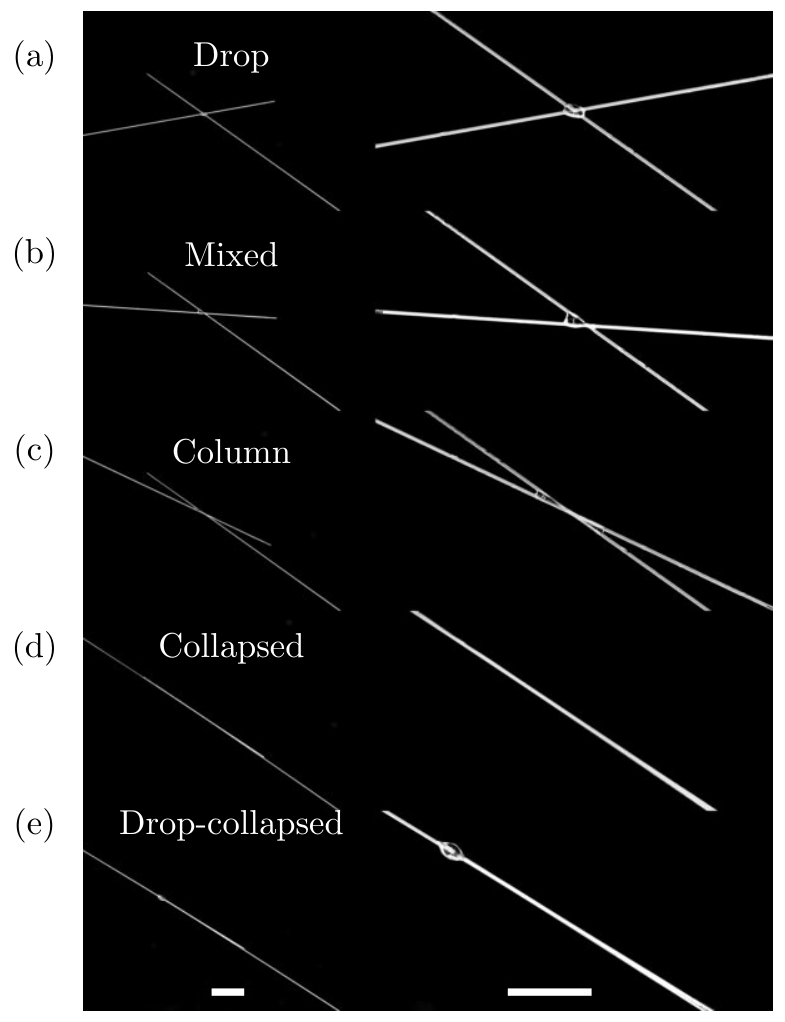}\\
\includegraphics[width=9cm]{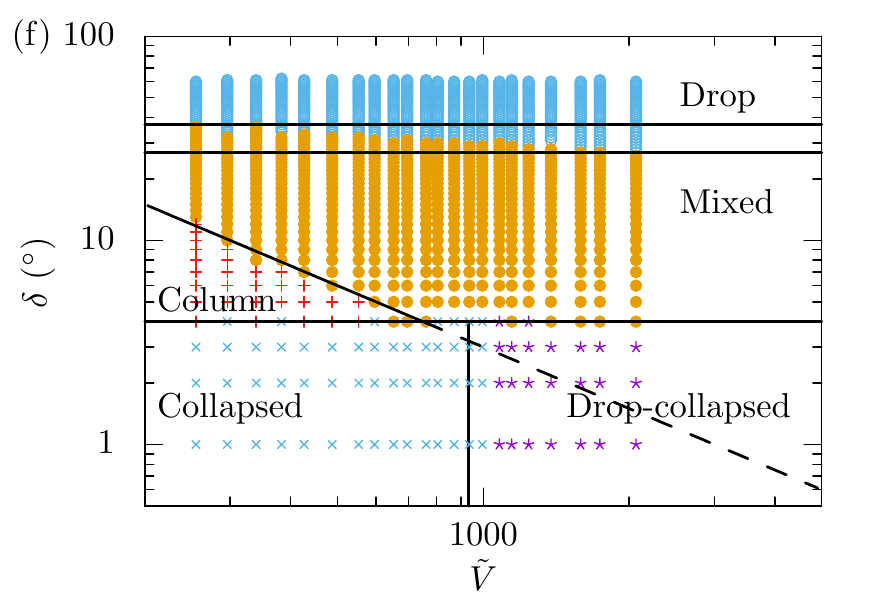}
\caption{Top views of the five different morphologies observed between two flexible fibers crossing as the angle $\delta$ is decreased: (a) drop, (b) mixed and (c) column morphology, as observed for rigid fibers. The flexibility of the fibers introduces two new morphologies at small tilting angle, the (d) collapsed and (e) drop-collapsed morphologies. The right column shows zoomed views of the morphologies.
    The scale bars represent $1$ cm.
    (f) Morphology diagram in the $(V/a^3,\delta)$ space for fibers of radius $a=125\,\mu{\rm m}$ showing the drop (blue circles), mixed (yellow circles), column (red crosses), collapsed (blue crosses) and drop-collapsed (purple stars) morphologies.
    The solid lines indicate the approximate transitions between the different regions.
    The dashed line represents the mixed-column transition that cannot be observed for flexible fibers.
    In this figure, $L_{free}=18.6$ mm and $L_{fib}=86.7$ mm.
    }\label{fig:morphology}
    \end{center}
\end{figure}

\section{Morphologies and transitions}
\subsection{Morphology diagram}

Our experimental observations highlight that the three liquid morphologies observed on rigid crossing fiber are recovered with flexible fibers as illustrated in Fig. \ref{fig:morphology}(a)-(c) \cite{Sauret:2015ch}: (i) the drop state for large angle (typically $\delta > 35^{\rm o}$) where the liquid collects at the node of the fiber and form a single hemispherical drop, (ii) a mixed morphology when the tilting angle is decreased and in which the liquid partially spreads into a column but a drop remains located on one side of the column (note that the side of the column where the drop is located is random and induced by perturbations present in the surrounding environment or introduced when changing the tilting angle) and (iii) a column state at small angle and small volume where the liquid spreads between the fibers and where the height of the column is typically comparable to the fiber diameter.

Considering crossed flexible fibers that are clamped on one end and free to move on their other end leads to two previously unreported morphologies, shown in Fig. \ref{fig:morphology}(d)-(e). This response is indeed caused by the flexibility of the crossed fibers. Indeed, the presence of the wetting liquid located at the node, {\textit{i.e.}}, the point where the two fibers cross, results in capillary forces that tend to pull on fibers and cause them to deflect inwards. This elastocapillary effect decreases the local distance between the fibers and the liquid can thus continue to spread spontaneously along the fibers. The magnitude of the capillary force increases, further bending the fibers, until they stick to one another, {\textit{i.e.}}, collapse, and the liquid can form a thin long column. The interfacial and bending energies therefore lead to the observed {\textit{collapsed}} morphology reported in Fig. \ref{fig:morphology}(d).
Finally, if more liquid is added to the system, the thin column keeps spreading until it reaches the full length of the sticking region, $2 L_{free}$, and a drop grows at one end of the column. This results in the {\textit{drop-collapsed}} morphology shown in Fig. \ref{fig:morphology}(e).

We experimentally characterize the region of existence of these five different morphologies observed for crossed flexible fibers having a constant total length $L_{fib}$ and a free length  $L_{free}$. We perform systematic experiments varying the liquid volume $\tilde{V}=V/a^3$ and the tilting angle $\delta$. We find five different final morphologies, as described above, and report our results in a ($\tilde{V}$, $\delta$) morphology diagram in Fig. \ref{fig:morphology}(f). Both the collapsed and drop-collapsed morphologies are observed below a tilting angle $\delta$ typically smaller than $5^{\rm o}$ for the fibers considered. Above this angle, we recover the morphology diagram of the rigid fibers as the capillary force is not sufficient to significantly bend the fibers at large tilting angle $\delta$. Indeed, the column/mixed morphology transition described by Sauret \textit{et al.} \cite{Sauret:2015ch} for rigid fibers satisfactorily captures the column/mixed morphology transition for flexible fibers as reported in Fig. \ref{fig:morphology}(f). In addition, the column/drop and mixed/drop morphologies transitions are also observed for a critical angle $\delta=35\,\pm\,5^{\rm o}$, as reported for rigid fibers. The key difference is therefore that, below a critical angle, the classical column morphology is not observed but instead the fibers stick to one another in a collapsed state. In addition, in this regime of small angles, we observe a state of partial spreading  for sufficiently large volume of liquid: a smaller drop remains at the edge of the collapsed region.

\subsection{Transitions at large angles: drop, mixed and column morphologies}

If the fibers are sufficiently rigid or the capillary force too weak to bend the fibers, the local angle at the node must be equal to the global tilting angle, that is $\delta=\phi$.
We measure the local angle $\phi$ as a function of the global angle $\delta$ in our system.
These results indicate that the capillary force is not sufficient to significantly bend the fibers for $\delta > 4^\circ$ for the fibers used in our experiments, which is the angle at which the collapse transition occurs. Note that the value at which the bending of the fibers becomes important depends on the flexibility of the fibers and thus on the elastocapillary length compared to the fibers length.
Therefore, the transition between the different regimes can be described by the model developed for rigid fibers, and described in previous publications \cite{Sauret:2014fb,Sauret:2015ch}.

\subsection{Transitions to the collapsed states}
The transition between the column or mixed morphology (Fig. \ref{fig:morphology}(b)-(c)) and the collapsed state (Fig. \ref{fig:morphology}(d)-(e)) is more complex. The bending and collapse of the fibers results from an elastocapillary equilibrium. To rationalize these results, despite the complex shape of the liquid column, we can estimate the energy associated with the bending of a fiber in the collapsed state and the energy associated to the capillary bridge induced by the liquid column in the collapsed state. Different simplified situations have considered this elastocapillary equilibrium between parallel sheets or fibers \cite{Bico:2004vi,Py2007a,Duprat:2012kya}. We denote $d$ the distance over which the fiber needs to be deflected for the fibers to stick to one another along $L_{free}$. Using geometrical considerations, we have

\begin{equation} \label{eq:geometry}
\sin \delta=\frac{d}{L_{free}}.
\end{equation}

Because the collapse happens in the limit of small angles, $\delta \ll 1$, we can simplify (\ref{eq:geometry}) to $d=\delta\,L_{free}$. Using the same notations as Bico {\textit{et al.}} \cite{Bico:2004vi}, we define the length of the wet segment of the fiber as $L_{wet}=L_{fib}-L_{dry}$, where $L_{fib}$ and $L_{dry}$ are the total length of a fiber and the dry part, respectively, the bending energy $E_c$ associated to the deformation of the two fibers that coalesce at a distance $L_{dry}$ is
\begin{equation}
E_c \simeq \frac{6\,B\,d^2}{{L_{dry}}^3}.
\end{equation}
To simplify, we consider the capillary force induced by a liquid column when the fiber are in the collapsed state and therefore the fibers are parallel. In this situation, the liquid column has a constant cross-sectional area $A \simeq \pi\,a^2$ on the entire free length $L_{wet}=2\,(L_{fib}-L_{dry})$. Therefore, the energy associated to the capillary force is
\begin{equation}
E_s \sim -2\,\gamma\,a\,(\pi-2)\,(L_{fib}-L_{dry}),
\end{equation}
where we assume that we can neglect the contribution of the terminal menisci and that the column is flat. This equation leads to an expression for $L_{dry}$ \cite{Duprat:2012kya}:
\begin{equation}
L_{dry} \simeq \left(\frac{9\,B\,{d}^2}{\gamma\,a\,(\pi-2)}\right)^{1/4}.
\end{equation}

Using equation (2), and considering that the fibers collapse when $L_{wet}=2\,L_{free}=2\,(L_{fib}-L_{dry})$ we obtain the condition on the tilting angle $\delta$
\begin{equation}\label{eq:delta_L}
\delta \simeq \frac{(L_{fib}-L_{free})^2}{L_{free}}\,\sqrt{\frac{\gamma\,a\,(\pi-2)}{9\,B}}.
\end{equation}

Introducing the elastocapillary length $\mathlarger{\ell}_{EC}$ defined in equation (\ref{bico}), and the ratio between the free length and the total length of the fiber $\alpha=L_{free}/L_{fib}$, this expression can be rewritten as:
\begin{equation}\label{eq:delta_L}
\delta \simeq \frac{L_{fib}}{\mathlarger{\ell}_{EC}}\, \frac{(1-\alpha)^2}{\alpha}\,\sqrt{\frac{(\pi-2)}{9}},
\end{equation}
which shows that increasing the total length of the fiber or decreasing the elastocapillary length leads to a collapsed state for larger tilting angle. This expression is valid for moderately flexible fibers, typically for $L_{fib}<\ell_{EC}$.

\begin{figure}
\begin{center}
\includegraphics[width=9cm]{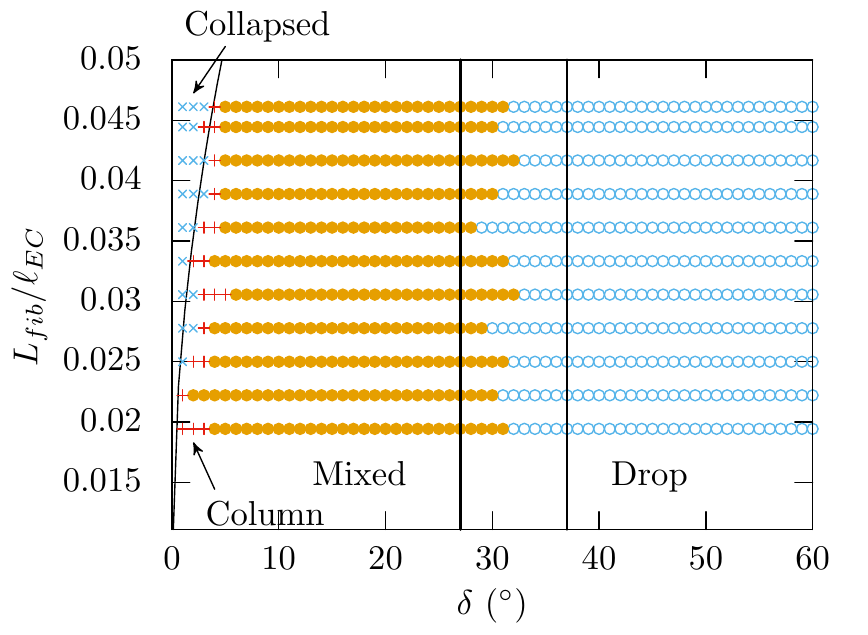}
  \end{center}
\caption{Morphologies observed for fibers of radius $a=125\,\mu{\rm m}$, $\ell_{EC} =1.8\,{\rm m}$ and varying $L_{fib}/\ell_{EC}$. $L_{free}=14\,{\rm mm}$, the volume $\tilde{V}=768$ and $a=125\,\mu{\rm m}$ are constant.
    The black solid curve represents equation (\ref{eq:delta_L}).
    The vertical black lines show the drop-mixed transition represented in Fig. \ref{fig:morphology}(f).
    }
\label{fig:transitionLfib}
\end{figure}

The experimental results obtained by varying the total length of the fiber $L_{fib}$ are in qualitative agreement with this expression as illustrated in Fig. \ref{fig:transitionLfib}: the longer the fibers, the easier is the coalescence of the fibers. Note that the order of magnitude estimate given here could be refined by considering the exact length, and shape of the column for a given volume. As a result, the energy associated to the capillary bridge between the fibers could be calculated between two crossed fibers \cite{Sauret:2015ch}. However, such calculations would require to know exactly the inter-fibers distance, which is modified by the elastocapillary equilibrium even before the collapsed state. 

With typical experimental values, $L_{fib}=87\,{\rm mm}$, $L_{free}=18.5\,{\rm mm}$, $B=EI=8.1\times 10^{-6}\,{\rm Pa \cdot m^4}$, $a=125\,{\rm mm}$, $\gamma=19.7\,{\rm mN \cdot m^{-1}}$, from (\ref{eq:delta_L}) we obtain $\delta \sim 3^{\rm o}$. This result is satisfactorily close to the observed value and shows that elastocapillary effects are indeed responsible for the collapse of the fibers at sufficiently small angle. However, we shall see that because of our assumption of a flat shape, the transition between the column or mixed morphology and the collapsed state is only an order of magnitude. In addition, because of the shape of the capillary bridge in the column and the collapsed morphologies, the angle necessary for the fibers to release from the sticking state will be larger than the angle for the fibers to collapse. Indeed, flexible fibers that collapse owing to capillary effects present a strong hysteresis as described in section 4.

\subsection{Transition between collapsed and drop-collapsed morphologies}
We now turn to the transition between the collapsed and drop-collapsed morphologies. Similarly to the situations described for parallel fibers, when the fibers collapse, the liquid shape can be assumed to be a long straight column with a cross-sectional area $A \simeq \pi\,a^2$ \cite{Duprat:2012kya}. Because the liquid can spread over a distance $L_{free}$ on each side of the node, the critical volume that describes the transition between collapsed and drop-collapsed morphologies is $V_c=2\,\pi\,a^2\,L_{free}$ leading to the critical dimensionless volume
\begin{equation} \label{eq_1}
\tilde{V}_c=\frac{2\,\pi\,L_{free}}{a}.
\end{equation}

\begin{figure}
    \centering
\includegraphics[width=9cm]{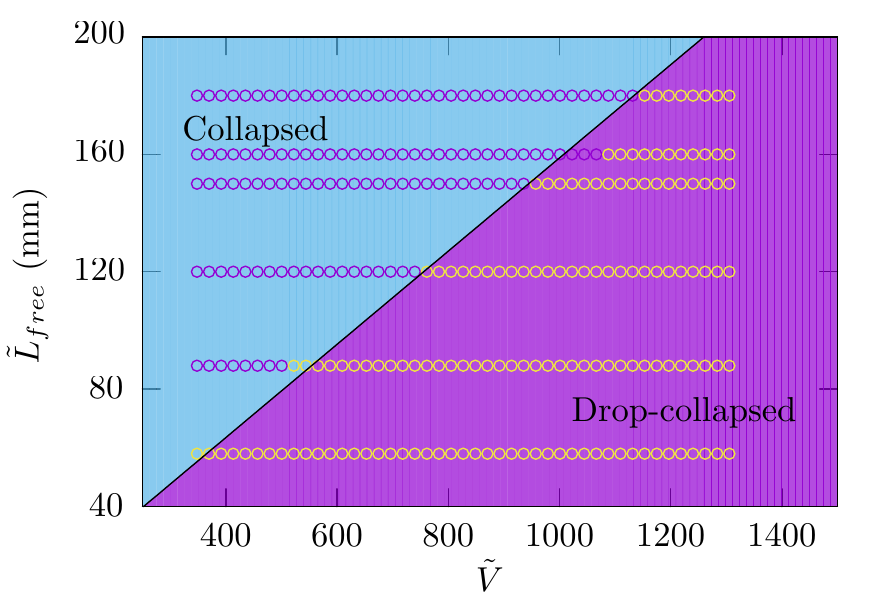}
\caption{Threshold value of the volume of liquid delimiting the collapsed and drop-collapsed states for varying length of the free end of the fiber $\tilde{L}_{free}=L_{free}/a$. The continuous black line is defined by equation (\ref{eq_1}) and the experimental parameters kept constant are $a=125\,\mu{\rm m}$ and $L_{fib}=110\,{\rm mm}$.}\label{fig:collapsed}
\end{figure}

This expression is in agreement with our experimental observation reported on the morphology diagram in Fig. \ref{fig:morphology}(f). In addition, we perform systematic experiments varying the free length of the fiber beyond the node, $L_{free}$. The volume of liquid at the transition is thus measured and the results shown in Fig. \ref{fig:collapsed} are well captured by equation (\ref{eq_1}). Beyond the threshold value $V_c$, adding more liquid leads to the formation of a liquid drop at one end of the thin column.

\section{Hysteresis}

\begin{figure}
\begin{center}
\includegraphics[width=8.5cm]{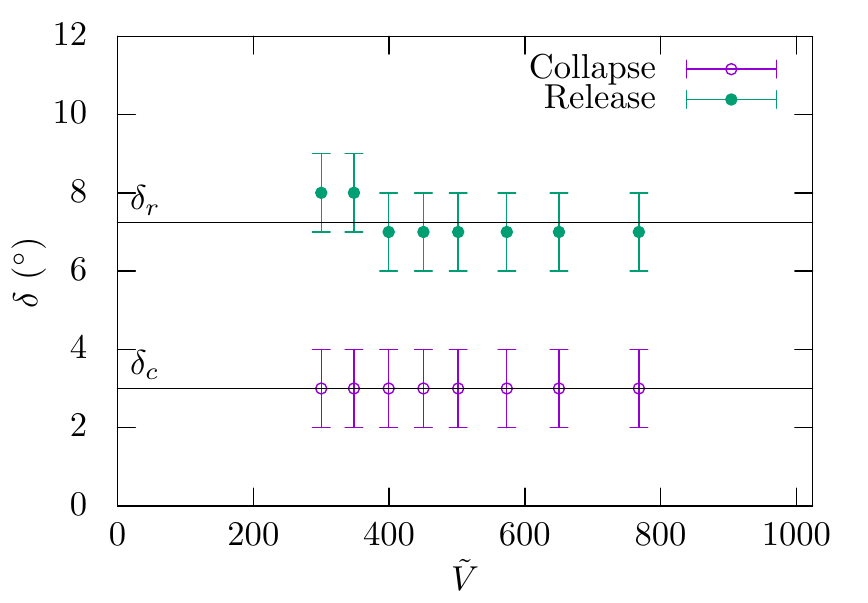}
\end{center}
    \caption{Critical angle for the fibers to collapse $\delta_c$ (hollow circles) when decreasing the tilting angle and to separate $\delta_r$ (filled circles) when the tilting angle is then increased for varying volume of liquid $V$. The fibers have a radius $a=125\,\mu{\rm m}$, a total length $L_{fib} =87\,{\rm mm}$ and a free length $L_{free} =18.6\,{\rm mm}$.}
\label{fig:hysteresis}
\end{figure}

The morphology diagram (Fig. \ref{fig:morphology}(f)) presented in this article is established by starting at large tilting angle $\delta$ and then decreasing the angle for each dimensionless volume of liquid $\tilde{V}$ considered. However, we also perform experiments considering the opposite situation: we start with collapsed fibers and incrementally increase the tilting angle until the collapsed to column/mixed morphology transition occurs (Fig. \ref{fig:morphology}(d)-(e) then Fig. \ref{fig:morphology}(b)-(c)). We observe that the critical tilting angle $\delta_r$ at which the fibers separate is much larger than the angle $\delta_c$ at which the fibers collapse. We have reported these hysteretic measurements in Fig. \ref{fig:hysteresis} for different volumes of liquid.

We observe that the amount of liquid seems to have no influence on the coalescence and separations angles in the range of volume considered. Indeed, when the fibers are collapsed, the liquid spreads along a thin liquid column and, at the volume considered, the volume is sufficient to reach the free end of the fibers. Therefore, the energy associated with the capillary force can be assumed constant as well as the bending energy that is necessary for the fibers to be released $\delta_r$.

Finally, we explore the effects of the free length $L_{free}$ on the critical angle $\delta_r$ for the separation of collapsed fibers. Indeed, the capillary force associated to the wetting of parallel fibers on a distance $L_{free}$ is $E_s=-2\,\gamma\,a\,(\pi-2)\,L_{free}$. Therefore, increasing the free length of the fibers qualitatively leads to an increase of the adhesion force, increasing the critical angle for the separation of the collapsed fibers. This hysteresis effect between the column/mixed and collapsed morphologies may modify significantly the lower part of the morphology diagram shown in Fig. \ref{fig:morphology}(f).

\section{Conclusion}

In this paper, we have experimentally illustrated that flexible crossed fibers share similar morphologies to the system of rigid fibers. In particular, at large tilting angle $\delta$, the three morphologies obtained for rigid crossed fibers are recovered. However, when decreasing the tilting angle, the energy required to collapse the fibers decreases while the capillary force increases. The capillary force first leads to a larger deflection of the fibers, namely the tilting angle between the fibers is locally decreased ($\phi < \delta$) slightly modifying the column to mixed morphology transition. In addition, we have reported two new morphologies at small tilting angle $\delta$: a column collapsed and a drop-collapsed morphology where the fibers are locally parallel. The transition between these two morphologies can be explained by considering the maximum amount of liquid that can spread into a long thin column around collapsed fibers. The transition between the column or mixed morphology and the collapsed morphologies can be captured by considering an elastocapillary equilibrium, as observed for parallel lamellae or fibers. We also observe a strong hysteresis in the system between the column/mixed and collapsed morphologies. {The experiments in this study are performed in the regime where $L_{fib}/\ell_{EC} \ll 1$. Future works will need to investigate the situation $L_{fib}/\ell_{EC} > 1 $ where the effects of the flexibility are expected to be even greater}. For instance, the influence of the flexibility of the fibers can be especially important in systems where the porosity of the fiber network need to be large as capillary effects will tend to align fibers and form large clusters \cite{Pokroy:2009ip,Bico:2004vi,Singh:2014bk}.

\section{Acknowledgments}

The work of AS is partially supported by a CNRS PICS grant n$^{\rm o}$07242. KS and ED acknowledge financial support from NYU Tandon School of Engineering's Undergraduate Summer Research Program. FB acknowledges that the research leading to these results partially received funding from the People Programme (Marie Curie Actions) of the European UnionÕs Seventh Framework Programme (FP7/2007-2013) under REA grant agreement 623541.

{\small \bibliography{SoftMatter_Elastocap_Fibers} 

\begin{thebibliography}{10}

\bibitem{FISHER1981528}
L.~R. Fisher and J.~N. Israelachvili, ``Experimental studies on the
  applicability of the kelvin equation to highly curved concave menisci,'' {\em
  Journal of Colloid and Interface Science}, vol.~80, no.~2, pp.~528 -- 541,
  1981.

\bibitem{Mastrangelo:1993uf}
C.~H. Mastrangelo and C.~H. Hsu, ``{Mechanical stability and adhesion of
  microstructures under capillary forces. I. Basic theory},'' {\em Systems},
  vol.~2, no.~1, pp.~33--43, 1993.

\bibitem{Willett:2000ic}
C.~D. Willett, M.~J. Adams, S.~A. Johnson, and J.~Seville, ``{Capillary bridges
  between two spherical bodies},'' {\em Langmuir}, pp.~9396--9405, 2000.

\bibitem{Roman:2010ipa}
B.~Roman and J.~Bico, ``{Elasto-capillarity: deforming an elastic structure
  with a liquid droplet},'' {\em Journal of Physics: Condensed Matter},
  vol.~22, pp.~493101--16, Nov. 2010.

\bibitem{Liu:2012gk}
J.~L. Liu and X.~Q. Feng, ``{On elastocapillarity: A review},'' {\em Acta
  Mechanica Sinica}, pp.~928--940, 2012.

\bibitem{Bico:2004vi}
J.~Bico, B.~Roman, L.~Moulin, and A.~Boudaoud, ``Adhesion: Elastocapillary
  coalescence in wet hair,'' {\em Nature}, vol.~432, no.~7018, pp.~690--690,
  2004.

\bibitem{Singh:2014bk}
K.~Singh, J.~R. Lister, and D.~Vella, ``{A fluid-mechanical model of
  elastocapillary coalescence},'' {\em Journal of Fluid Mechanics}, vol.~745,
  pp.~621--646, 2014.

\bibitem{Pokroy:2009ip}
B.~Pokroy, S.~H. Kang, L.~Mahadevan, and J.~Aizenberg, ``{Self-organization of
  a mesoscale bristle into ordered, hierarchical helical assemblies},'' {\em
  Science}, vol.~323, no.~5911, pp.~237--240, 2009.

\bibitem{Boudaoud:2007hf}
A.~Boudaoud, J.~Bico, and B.~Roman, ``{Elastocapillary coalescence: aggregation
  and fragmentation with a maximal size},'' {\em Physical Review E}, vol.~76,
  no.~6, p.~060102, 2007.

\bibitem{Wang:2014ej}
Q.~Wang, B.~Su, H.~Liu, and L.~Jiang, ``{Chinese brushes: Controllable liquid
  transfer in ratchet conical hairs},'' {\em Advanced Materials}, vol.~26,
  no.~28, pp.~4889--4894, 2014.

\bibitem{Chandra:2009jc}
D.~Chandra, S.~Yang, and A.~A. Soshinsky, ``{Biomimetic ultrathin whitening by
  capillary-force-induced random clustering of hydrogel micropillar arrays},''
  {\em ACS Applied Materials}, vol.~1, no.~8, pp.~1698--1704, 2009.

\bibitem{Chandra:2010eo}
D.~Chandra and S.~Yang, ``{Stability of high-aspect-ratio micropillar arrays
  against adhesive and capillary forces},'' {\em Accounts of Chemical
  Research}, vol.~43, no.~8, pp.~1080--1091, 2010.

\bibitem{DeVolder:2013cs}
M.~De~Volder and A.~J. Hart, ``{Engineering Hierarchical Nanostructures by
  Elastocapillary Self-Assembly},'' {\em Angewandte Chemie International},
  vol.~52, no.~9, pp.~2412--2425, 2013.

\bibitem{Chakrapani:2004up}
N.~Chakrapani, B.~Wei, and A.~Carrillo, ``{Capillarity-driven assembly of
  two-dimensional cellular carbon nanotube foams},'' {\em Proceedings of the
  National Academy of Sciences}, vol.~101, no.~12, pp.~4009--4012, 2004.

\bibitem{Hartung:1967tm}
R.~Clark, ``Impact of oil pollution on seabirds,'' {\em Environmental Pollution
  Series A, Ecological and Biological}, vol.~33, no.~1, pp.~1--22, 1984.

\bibitem{Py2007a}
C.~Py, R.~Bastien, J.~Bico, B.~Roman, and A.~Boudaoud, ``3d aggregation of wet
  fibers,'' {\em EPL}, vol.~77, no.~4, p.~44005, 2007.

\bibitem{Duprat:2012kya}
C.~Duprat, S.~Protiere, A.~Y. Beebe, and H.~A. Stone, ``Wetting of flexible
  fibre arrays,'' {\em Nature}, vol.~482, no.~7386, pp.~510--513, 2012.

\bibitem{elettro2016drop}
H.~Elettro, S.~Neukirch, F.~Vollrath, and A.~Antkowiak, ``In-drop capillary
  spooling of spider capture thread inspires hybrid fibers with mixed
  solid--liquid mechanical properties,'' {\em Proceedings of the National
  Academy of Sciences}, p.~201602451, 2016.

\bibitem{schulman2016elastocapillary}
R.~D. Schulman, A.~Porat, K.~Charlesworth, A.~Fortais, T.~Salez,
  E.~Rapha{\"e}l, and K.~Dalnoki-Veress, ``Elastocapillary bending of
  microfibers around liquid droplets,'' {\em arXiv preprint arXiv:1607.05990},
  2016.

\bibitem{Minor:1959wa}
F.~W. Minor, A.~M. Schwartz, E.~Wulkow, and L.~C. Buckles, ``The migration of
  liquids in textile assemblies part ii: The wicking of liquids in yams,'' {\em
  Textile Research Journal}, vol.~29, no.~12, pp.~931--939, 1959.

\bibitem{Miller:1967vy}
B.~Miller, A.~B. Coe, and P.~N. Ramachandran, ``{Liquid rise between filaments
  in a V-configuration},'' {\em Textile Research Journal}, vol.~37, no.~11,
  pp.~919--924, 1967.

\bibitem{Eadie:2011hm}
L.~Eadie and T.~K. Ghosh, ``Biomimicry in textiles: past, present and
  potential. an overview,'' {\em Journal of the Royal Society Interface},
  p.~rsif20100487, 2011.

\bibitem{Mullins:2004es}
B.~J. Mullins, I.~E. Agranovski, R.~D. Braddock, and C.~M. Ho, ``Effect of
  fiber orientation on fiber wetting processes,'' {\em Journal of Colloid and
  Interface Science}, vol.~269, no.~2, pp.~449 -- 458, 2004.

\bibitem{Contal:2004bg}
P.~Contal, J.~Simao, D.~Thomas, T.~Frising, and S.~Call{\'e}, ``{Clogging of
  fibre filters by submicron droplets. Phenomena and influence of operating
  conditions},'' {\em Journal of Aerosol}, vol.~35, no.~2, pp.~263--278, 2004.

\bibitem{Ju:2012gl}
J.~Ju, H.~Bai, Y.~Zheng, T.~Zhao, R.~Fang, and L.~Jiang, ``{A multi-structural
  and multi-functional integrated fog collection system in cactus},'' {\em
  Nature Communications}, vol.~3, p.~1247, 2012.

\bibitem{Park:2013go}
K.~C. Park, S.~S. Chhatre, S.~Srinivasan, and R.~E. Cohen, ``{Optimal design of
  permeable fiber network structures for fog harvesting},'' {\em Langmuir},
  vol.~29, no.~43, pp.~13269--13277, 2013.

\bibitem{LeBoeuf:2014hu}
R.~LeBoeuf and E.~de~la Jara, ``{Quantitative goals for large-scale fog
  collection projects as a sustainable freshwater resource in northern
  Chile},'' {\em Water International}, vol.~39, no.~4, pp.~431--450, 2014.

\bibitem{Sauret:2015ch}
A.~Sauret, F.~Boulogne, B.~Soh, E.~Dressaire, and H.~A. Stone, ``Wetting
  morphologies on randomly oriented fibers,'' {\em The European Physical
  Journal E}, vol.~38, no.~6, p.~62, 2015.

\bibitem{Bintein:2015tw}
P.~B. Bintein, {\em {Dynamiques de Gouttes Funambules: Applications {\`a} la
  Fabrication de Laine de Verre}}.
\newblock PhD thesis, Dynamiques de Gouttes Funambules: Applications {\`a} la
  Fabrication de Laine de Verre, Universit{\'e} Pierre et Marie Curie, Jan.
  2015.

\bibitem{Princen:1970ts}
H.~Princen, ``Capillary phenomena in assemblies of parallel cylinders: {III}.
  liquid columns between horizontal parallel cylinders,'' {\em Journal of
  Colloid and Interface Science}, vol.~34, no.~2, pp.~171 -- 184, 1970.

\bibitem{Carroll:1986uia}
B.~Carroll, ``Equilibrium conformations of liquid drops on thin cylinders under
  forces of capillarity. {A} theory for the roll-up process,'' {\em Langmuir},
  vol.~2, no.~2, pp.~248--250, 1986.

\bibitem{Quere:1999wv}
D.~Qu\'er\'e, ``Fluid coating on a fiber,'' {\em Annual Review of Fluid
  Mechanics}, vol.~31, no.~1, pp.~347--384, 1999.

\bibitem{Sutter:2010jf}
B.~Sutter, D.~B{\'e}mer, J.-C. Appert-Collin, D.~Thomas, and N.~Midoux,
  ``Evaporation of liquid semi-volatile aerosols collected on fibrous
  filters,'' {\em Aerosol Science and Technology}, vol.~44, no.~5,
  pp.~395--404, 2010.

\bibitem{Boulogne:2015kl}
F.~Boulogne, A.~Sauret, B.~Soh, E.~Dressaire, and H.~A. Stone, ``Mechanical
  tuning of the evaporation rate of liquid on crossed fibers,'' {\em Langmuir},
  vol.~31, no.~10, pp.~3094--3100, 2015.

\bibitem{Lorenceau:2004da}
E.~Lorenceau, C.~Clanet, and D.~Qu\'er\'e, ``Capturing drops with a thin
  fiber,'' {\em Journal of Colloid and Interface Science}, vol.~279, no.~1,
  pp.~192 -- 197, 2004.

\bibitem{Zhang:2015dt}
K.~Zhang, F.~Liu, A.~J. Williams, X.~Qu, J.~J. Feng, and C.~H. Chen,
  ``{Self-propelled droplet removal from hydrophobic fiber-based coalescers},''
  {\em Physical Review Letters}, vol.~115, no.~7, p.~074502, 2015.

\bibitem{gilet2009digital}
T.~Gilet, D.~Terwagne, and N.~Vandewalle, ``Digital microfluidics on a wire,''
  {\em Applied Physics Letters}, vol.~95, no.~1, p.~014106, 2009.

\bibitem{Gilet:2010ti}
T.~Gilet, D.~Terwagne, and N.~Vandewalle, ``Droplets sliding on fibres,'' {\em
  The European Physical Journal E}, vol.~31, pp.~253--262, 2010.
\newblock 10.1140/epje/i2010-10563-9.

\bibitem{Boulogne2012}
F.~Boulogne, L.~Pauchard, and F.~Giorgiutti-Dauphin\'{e}, ``Instability and
  morphology of polymer solutions coating a fibre,'' {\em Journal of Fluid
  Mechanics}, vol.~704, pp.~232--250, 7 2012.

\bibitem{weyer2015compound}
F.~Weyer, M.~Lismont, L.~Dreesen, and N.~Vandewalle, ``Compound droplet
  manipulations on fiber arrays,'' {\em Soft matter}, vol.~11, no.~36,
  pp.~7086--7091, 2015.

\bibitem{Blanchette:2016ds}
A.~Bick, F.~Boulogne, A.~Sauret, and H.~A. Stone, ``Tunable transport of drop
  on a vibrating fiber,'' {\em Applied Physics Letters}, vol.~107, no.~18,
  p.~181604, 2015.

\bibitem{Piroird:2009hu}
K.~Piroird, C.~Clanet, E.~Lorenceau, and D.~Qu\'er\'e, ``Drops impacting
  inclined fibers,'' {\em Journal of Colloid and Interface Science}, vol.~334,
  pp.~70--74, 2010.

\bibitem{Dressaire:2015hp}
E.~Dressaire, A.~Sauret, F.~Boulogne, and H.~A. Stone, ``Drop impact on a
  flexible fiber,'' {\em Soft Matter}, vol.~12, pp.~200--208, 2016.

\bibitem{Carroll:1976uj}
B.~Carroll, ``The accurate measurement of contact angle, phase contact areas,
  drop volume, and laplace excess pressure in drop-on-fiber systems,'' {\em
  Journal of Colloid and Interface Science}, vol.~57, no.~3, pp.~488 -- 495,
  1976.

\bibitem{McHale:2001vy}
G.~McHale, M.~I. Newton, and B.~J. Carroll, ``The shape and stability of small
  liquid drops on fibers,'' {\em Oil and Gas Science and Technology - Rev.
  IFP}, vol.~56, no.~1, pp.~47--54, 2001.

\bibitem{McHale:2002wu}
G.~McHale and M.~I. Newton, ``Global geometry and the equilibrium shapes of
  liquid drops on fibers,'' {\em Colloids and Surfaces A: Physicochemical and
  Engineering Aspects}, vol.~206, no.~1-3, pp.~79 -- 86, 2002.

\bibitem{Wu:2014eb}
X.~F. Wu, M.~Yu, Z.~Zhou, A.~Bedarkar, and Y.~Zhao, ``{Droplets engulfing on a
  filament},'' {\em Applied Surface Science}, vol.~294, pp.~49--57, 2014.

\bibitem{Sauret:2014fb}
A.~Sauret, A.~D. Bick, C.~Duprat, and H.~A. Stone, ``Wetting of crossed fibers:
  Multiple steady states and symmetry breaking,'' {\em EPL}, vol.~105, no.~5,
  p.~56006, 2014.

\bibitem{Sauret:2015ba}
A.~Sauret, F.~Boulogne, D.~C{\'e}bron, E.~Dressaire, and H.~A. Stone,
  ``{Wetting morphologies on an array of fibers of different radii},'' {\em
  Soft Matter}, vol.~11, pp.~4034--4040, May 2015.

\bibitem{soleimani2015capillary}
M.~Soleimani, R.~J. Hill, and T.~G.~M. van~de Ven, ``Capillary force between
  flexible filaments,'' {\em Langmuir}, vol.~31, no.~30, pp.~8328--8334, 2015.

\bibitem{Claussen2011}
J.~O. Claussen, {\em {Elasticity and Morphology of Wet Fibers}}.
\newblock PhD thesis, Elasticity and Morphology of Wet Fibers, University of
  G\"ottingen, 2011.

\end{thebibliography}
    \bibliographystyle{ieeetr}}

\end{document}